\documentclass[aps, preprint, 12pt]{article}
\usepackage{times}
\usepackage{amssymb}
\usepackage{amsmath}
\usepackage{array}
\usepackage{float}
\usepackage[margin=2cm]{geometry}
\usepackage{fancyhdr}
\usepackage{cite}
\usepackage[format=plain,=small,=onehalfspacing,labelfont=bf,labelsep=period]{caption}
\usepackage{authblk}

\usepackage{color}
\usepackage{pdfcolmk} % allow colors to span pages
\usepackage{setspace}
\usepackage{soul}
\usepackage[hidelinks]{hyperref}

\usepackage{pdfpages}    %need this package

\ifx\pdfoutput\undefined
\usepackage{graphicx}
\else
\usepackage{pdflscape} % needed for landscape orientation

\usepackage[version=3]{mhchem} %for typesetting chemical formulae

\newcolumntype{x}[1]{>{\raggedright\hspace{0pt}}p{#1}}

\graphicspath{{.},{./f/}}

\usepackage{threeparttable}
\title{\textbf{Creating a Microstructure Latent Space with Rich Material Information for Multiphase Alloy Design}}

\author{\normalsize{Xudong Ma$^{\text{a}}$, Yuqi Zhang$^{\text{a}}$, Chenchong Wang$^{\text{a},~\text{*}}$, Ming Wang$^{\text{b}}$, MingXin Huang$^{\text{b}}$, Wei Xu$^{\text{a},~\text{*}}$}}
\affil{$^\text{a}$ State Key Laboratory of Rolling and Automation, Northeastern University, Shenyang, Liaoning 110819, China}
\affil{$^\text{b}$ Department of Mechanical Engineering, The University of Hong Kong, Hong Kong, China}
\affil{$^\text{*}$ Corresponding author: wangchenchong@ral.neu.edu.cn; xuwei@ral.neu.edu.cn}

\renewcommand{\figurename}{Fig.}
\usepackage{indentfirst}
\usepackage{multirow}
\usepackage{makecell}
\usepackage{enumerate}
\usepackage{secdot}
\sectiondot{subsection}
\sectiondot{subsubsection}

\begin{document}\begin{sloppypar}

% show the title, author and date
\date{}
\maketitle
%%%%%%%%%%%%%%%%%%%%%%%%%%%%%%%%%%%%%%%%%%%%%%%%%%%%%%%%%%%%
%% Abstract %%%%%%%%%%%%%%%%%%%%%%%%%%%%%%%%%%%%%%%%%%%%%%%%
%%%%%%%%%%%%%%%%%%%%%%%%%%%%%%%%%%%%%%%%%%%%%%%%%%%%%%%%%%%%

\begin{abstract}
\noindent
The intricate microstructure serves as the cornerstone for the 
composition/processing-structure-property (CPSP) connection in multiphase alloys. Traditional alloy design methods often overlook microstructural details, which diminishes the reliability and effectiveness of the outcomes.  This study introduces an improved alloy design algorithm that integrates authentic microstructural information to establish precise CPSP relationships. The approach utilizes a deep-learning framework based on a variational autoencoder to map real microstructural data to a latent space, enabling the prediction of composition, processing steps, and material properties from the latent space vector. By integrating this deep learning model with a specific sampling strategy in the latent space, a novel, microstructure-centered algorithm for multiphase alloy design is developed. This algorithm is demonstrated through the design of a unified dual-phase steel, and the results are assessed at three performance levels. Moreover, an exploration into the latent vector space of the model highlights its seamless interpolation ability and its rich material information content. Notably, the current configuration of the latent space is particularly advantageous for alloy design, offering an exhaustive representation of microstructure, composition, processing, and property variations essential for multiphase alloys. \\
\\
{\singlespace \footnotesize \noindent \textbf{Keywords:} Microstructure; Latent space; Alloy design; Variational autoencoder.}
\end{abstract}

%%%% IMMI only have space for 4--6 keywords

\newpage

% Graphical abstract:
\noindent
{\large\textbf{Graphical abstract:}}
% \begin{center}\includegraphics[width=3.3in]{figures/graphical_abstract.pdf}\end{center}
\begin{center}\includegraphics[width=6.5in]{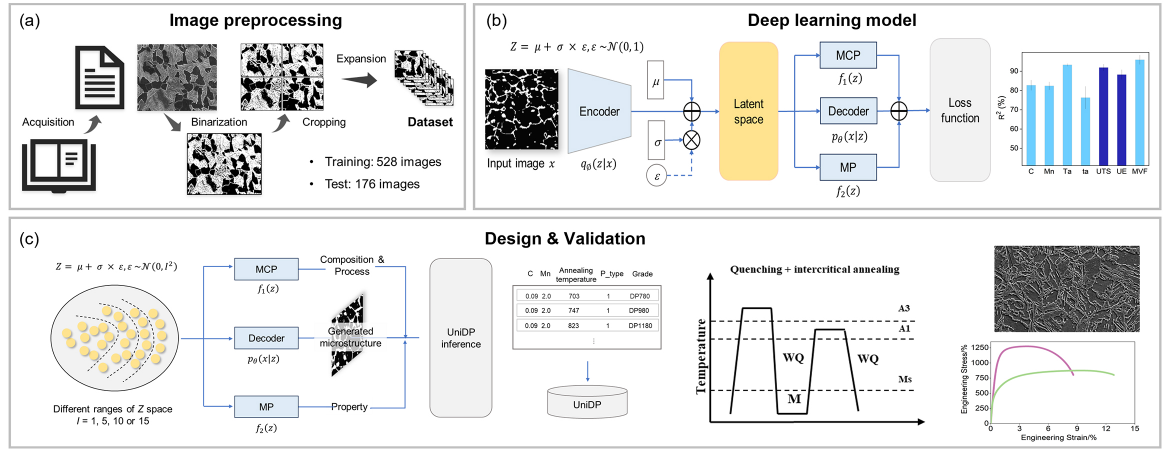}\end{center}

\newpage

\section{Introduction}
%%%%% Changes in the Machine Learning Paradigm

In the realm of alloy design, the paramount goal is to engineer novel materials suitable for targeted applications. Central to this is the elucidation of composition/processing–structure–property (CPSP) relationships. These relationships, traditionally derived from experimental data, such as the Hall–Petch relationship linking grain size to yield strength~\cite{RN282,RN406}, or the influence of finely dispersed second phases on strength augmentation~\cite{RN288}, underscore the critical role of microstructure. The complexity and heterogeneity of multiphase alloys, however, necessitate a refined understanding of these microstructural features. In such cases, conventional CPSP relationships often fall short, as they may not adequately reflect the nuanced and often nonlinear interactions within the microstructure of these alloys.

In contrast to conventional methods, machine learning (ML) techniques can be used to model complex and nonlinear relationships, which is fundamental for their applications in alloy design~\cite{RN310,RN344}. Previous studies~\cite{RN247, RN248} in this area have demonstrated that ML models that use the composition and process parameters of alloys as model inputs, can predict mechanical properties. These models were then combined with heuristic optimization algorithms to search for optimal solutions and develop novel alloys. Although these ML-based methods are fast and relatively simple~\cite{RN249}, they overlook the key role of the microstructure. To address this limitation, the incorporation of microstructural information is crucial. Some microstructural metrics, such as the volume fraction of a certain phase~\cite{RN175,RN173}, can be used as additional input to the ML model to improve the prediction accuracy of the mechanical properties. However, relying solely on simple metrics might be insufficient for alloy systems with complex microstructures.

Complex microstructures can often be represented as images using scanning electron microscopy. Hence, the use of deep learning methods to process microstructural images and extract key features is important for alloy design. Generative models, such as the generative adversarial network (GAN)~\cite{RN304} and variational autoencoder (VAE)~\cite{RN302}, offer new solutions for alloy design~\cite{RN187,RN188,RN217,RN177,RN178,RN179}. These models typically perform unsupervised learning on training images to output low-dimensional representations and link them to the compositions, processes, or properties relevant to alloy design. For instance, Cao et al.~\cite{RN177} applied a conditional GAN to extract the microstructural features of Ti–6Al–4V and established a relationship between the process and microstructure. However, their model required process parameters and random vectors as inputs to generate the microstructure, making it impossible to achieve an inverse processing design based on the microstructure. In addition, GAN models are associated with challenges during training, including pattern collapse~\cite{RN243}. Therefore, building a generative model with a stable training process and determining suitable input and output features are critical aspects of alloy design. Kusampudi et al.~\cite{RN178} applied the VAE to extract descriptors from the microstructure of synthetic dual-phase (DP) steels, built relationships between the descriptors and properties, and used Bayesian optimization to determine the best combination of descriptors. Similarly, Kim et al.~\cite{RN179} employed the VAE and Gaussian process regression to determine the optimal microstructure. These methods use microstructural images as inputs to avoid subjectivity when selecting the microstructural features. However, the relationship between the composition/process and the microstructure is missing in these approaches, as they were developed using synthetic microstructures. Consequently, designed microstructures with ideal properties may not be realized experimentally.

A unified DP (UniDP) steel is a type of unified steel~\cite{RN224} that aims to meet different performance requirements with a single-alloy composition and different process parameters. The use of this type of steel can significantly simplify the systematic issues encountered in automotive steel production and recycling. The microstructures of UniDP steels are mainly ferrite and martensite, and an appropriate balance between them is crucial for the success of the UniDP steel design. However, the complexity and diversity of microstructural morphology pose challenges in this regard. Hence, a microstructure-centered UniDP steel design is required.

 In this study, we proposed a novel deep-learning algorithm specifically tailored for the design of multiphase alloys, emphasizing the pivotal role of microstructural characteristics in the alloy design process. Through VAE, we transformed microstructural images of dual-phase steels into a compact representation within a latent space, capturing the complex microstructural features of alloys. This latent space is then correlated with the alloy's composition, processing parameters, and mechanical properties to establish complete CPSP relationships. To emphasize the usefulness of the algorithm in the field of the design of multiphase alloy, we proposed a microstructure-centered design method for UniDP steels based on the principles of physical metallurgy (PM) and a specific sampling strategy within the latent space. Distinct from previous methods, it leverages experimental microstructures and complete CPSP connections, accelerating the design of UniDP steels. In addition, a visualization of the latent space demonstrated that integrating authentic microstructural details with precise CPSP linkages results in a continuously interpolated and information-rich mapping space. This space provides a robust foundation for the effective design and discovery of novel multiphase alloys, emphasizing the pivotal role of microstructural features in advancing the frontier of alloy design.

\section{Methodology}

%%%%% Generalized Segmentation Method for Alloy Images
\subsection{Microstructure-centered alloy design framework}
\begin{figure}[H]
\centering

\includegraphics[width=6.5in]{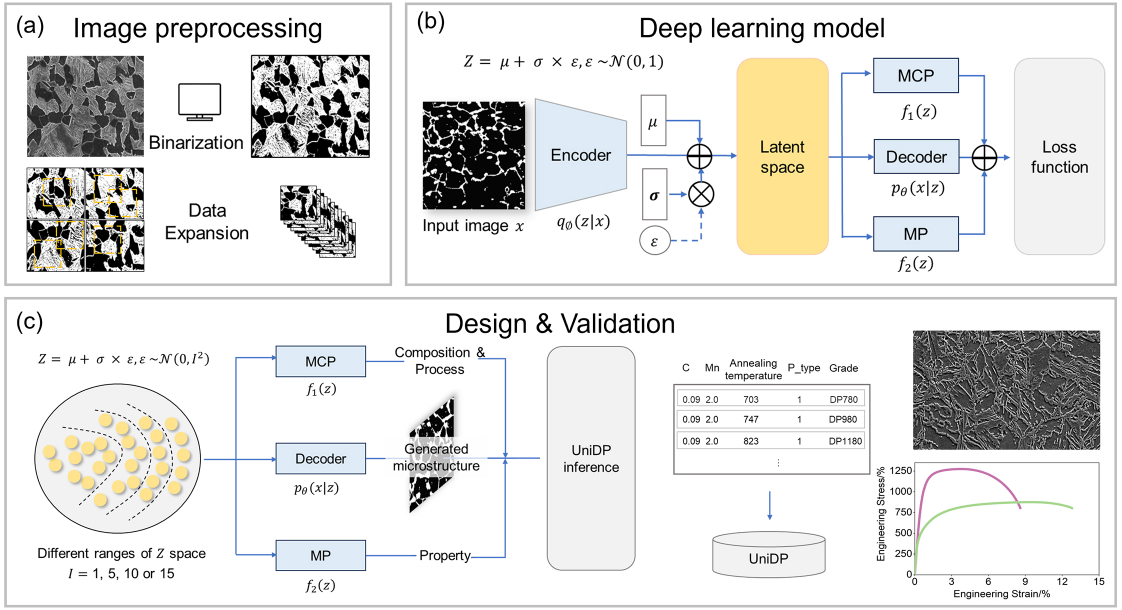}
\caption{\textbf{Microstructure-centered alloy design framework.} (a) Preprocessing of literature-sourced microstructural images through binarization and data expansion. (b) Construction of a deep learning model comprising an encoder, a decoder, and dual multilayer perceptrons to establish the microstructure-centered composition/processing–structure–property (CPSP) connections. (c) Design and validation process involving random sampling within the latent vector space, guided by physical metallurgy principles, to identify potential unified dual-phase (UniDP) steel candidates.}
\label{fig:framework}
\end{figure}

Our framework for alloy design is fundamentally rooted in the microstructure and reveals three distinct phases, as shown in Fig.~\ref{fig:framework}. First, we obtained microstructural images from prior studies~\cite{RN191,RN192,RN193,RN194}. These images were subjected to binarization and data augmentation to enhance their quality and variability, providing the basis for accurate modeling. The core of our framework is a deep-learning architecture designed to establish robust CPSP connections. This architecture comprises a VAE network~\cite{RN302} with an encoder and a decoder supplemented by dual multilayer perceptrons (MLPs). Together they work to predict the composition, processing parameters, and properties of the alloy directly from an initial microstructural image. The final phase involved applying the CPSP relationships derived from our modeling for alloy design. We explored the latent vector space created by our deep learning model to identify promising alloy candidates. Our selection process was guided by the foundational principles of PM, which ensured that our choices were grounded in scientific reasoning. To validate our approach, experiments were conducted to test the properties of the designed UniDP steels.

In the initial phase, the creation of a dataset was crucial. Given the high costs associated with alloy production and testing, creating a dataset that accurately mirrors the structural and performance characteristics of DP steels is challenging. To overcome this limitation, we collected data on DP steels from highly cited papers~\cite{RN191,RN192,RN193,RN194}, including the chemical composition, heat treatment parameters, scanning electron microscope (SEM) images, and mechanical properties. We collected and preprocessed 22 samples (Supplementary Table~\ref{output distribution} and Supplementary Fig.~\ref{fig: details of data}). The heat treatment processes were divided into three distinct groups for ease of prediction. The microstructural images were converted into a binary format (with black representing ferrite and white representing martensite) based on the martensite volume fraction (MVF) reported in literature. Then, these images were normalized, cropped, and expanded to facilitate the incorporation of microstructural data into the model training phase and prevent model overfitting. Details of image preprocessing are shown in \textbf{Section 2.2.1.}

The second phase involved the development of a deep learning model. To overcome the lack of generalization in microstructural feature selection across various alloy systems, we devised a VAE-centric deep learning model (VAE–DLM) to establish a robust and comprehensive CPSP linkage. The key innovation of the model is the ability of the VAE to autonomously extract authentic microstructural features through iterative optimization, forming a latent space~\cite{RN259,RN260}. Meanwhile, the two MLPs, namely the MCP and MP models, bolster the representational capacity of the VAE by imposing regularization constraints derived from the composition, process, and property information of the material. The VAE–DLM was then trained using the DP steel dataset constructed in the initial phase.

In the final stage, the design process leveraged the CPSP relationships established in the second phase by probing the latent vector space of the VAE–DLM. This exploration yielded an array of latent vectors, each of which could be utilized by the VAE–DLM to predict the corresponding compositions, processes, properties, and microstructural images. The fundamental principles of PM provided guidance for the identification of potential UniDP steel candidates. Experiments were conducted to validate the design outcomes and confirm the feasibility of the microstructure-centered alloy design approach.

%%%%% 
\subsection{Details of the design framework}
\subsubsection{Preprocessing of microstructural images of literature data}

The raw image data were composed of microstructural images of 22 steels with different chemical compositions. Initially, we cropped the images to eliminate extraneous parts (e.g., markers) and binarized them based on the MVF reported in literature. Then, each image was divided into four equally sized sub-images and all the data were partitioned into four groups, with each group containing one sub-image from each original image. To mitigate the overfitting problem, we applied an offline data augmentation technique: we randomly sampled each sub-image eight times at a pixel size of 224 × 224. Ultimately, we obtained 176 sub-images for each group, totaling 704 sub-images. 

\subsubsection{Ultimate tensile strength and uniform elongation criteria of UniDP}

In general, the commercial performance standards for DP780, DP980, and DP1180 require ultimate tensile strength (UTS) values exceeding 780, 980, and 1180 MPa, respectively, along with corresponding total elongation (TEL) values exceeding 12$\%$, 8$\%$, and 5$\%$~\cite{RN265}. However, the tensile plate specimen size used in the commercial performance standard (No. 5, JIS Z 2241 standard) differed from the plate specimen size used in this study, which are A25 (25 mm in length and 6 mm in width). Additionally, since we used UTS and uniform elongation (UE) for evaluation during the design process of UniDP steel, the commercial performance standard lacked information about UE. Therefore, we needed to perform a standard conversion to obtain the TEL standard under A25 conditions. Then, we collected complete data for commercial DP780, DP980, and DP1180, adjusted their TEL values to equivalent values under A25 conditions, and calculated the adjusted TEL values and corresponding UE ratios. We averaged the ratios for the three steel grades. Finally, we divided the TEL criteria under A25 conditions by this average to determine the final UE criteria. 

The TEL criteria for DP780, DP980, and DP1180 were 11.90$\%$, 7.93$\%$, and 4.96$\%$, respectively, under A25 conditions (calculated using equations from the literature~\cite{RN264}). The ratios of the adjusted TEL values (computed using the same formulas~\cite{RN264}) to the corresponding UE values for the collected commercial DP780, DP980, and DP1180 were 1.63, 1.91, and 1.42, respectively, resulting in an average ratio of 1.65. Finally, the UE criteria for DP780, DP980, and DP1180 were 7.21$\%$, 4.81$\%$, and 3.01$\%$, respectively.

\subsubsection{Evaluation metrics for the prediction}

We used two metrics to assess the predictive ability of the model: squared correlation coefficient ($R^{2}$) and mean absolute error ($MAE$). These metrics can be calculated as follows:

\begin{equation}
R^{2} = \frac{\left ( n\sum_{n}^{i=1} f\left (x_{i}   \right ) y_{i}- \sum_{n}^{i=1} f\left (x_{i}   \right ) \sum_{n}^{i=1} y_{i}   \right)^{2}}{\left ( n\sum_{n}^{i=1} f\left (x_{i}   \right )^{2}-\left (  \sum_{n}^{i=1} f\left (x_{i}   \right )\right )^{2}   \right ) \left (  n\sum_{n}^{i=1} y_{i}^{2}-\left (  \sum_{n}^{i=1} y_{i}\right )^{2}\right )}
\end{equation}

\begin{equation}
MAE = \frac{1}{n}{\textstyle \sum_{n}^{i=1}}\left | f\left (x_{i} \right ) - y_{i} \right | 
\end{equation}

\noindent
where $n$ is the number of samples and $f\left (x_{i} \right )$ and $y_{i}$  represent the predicted and experimental values of the $i_{th}$ sample, respectively.

Given the limited data, we employed 4-fold cross-validation to evaluate the predictive performance of the model, which entailed using one group for testing and the remainder for training; this process was repeated four times, each with a different test group.

\subsection{Experimental validation}
\subsubsection{Experimental validation of designed steel}

The alloy with the designed chemical composition was smelted and forged into a steel ingot weighing approximately 50 kg, hot-rolled into slabs with a thickness of 3 mm, and then cooled in a furnace. Following the removal of the oxide layer by pickling, the slabs were cold-rolled to produce sheets with a thickness of 1.4 mm. The sheets were then heated to 900 $ \rm^\circ C $ for a duration of 5 min and quenched. Intercritical annealing regimes with varying temperatures and durations were subsequently employed to achieve the desired properties and microstructures (see Supplementary Table~\ref{experimental alloys}). Finally, the specimens were subjected to SEM characterization and performance testing using a JSM-7800F field-emission scanning electron microscope and a tensile testing machine, respectively. The tensile specimens, fabricated in alignment with the rolling direction of the plates, were prepared in accordance with the ASTM E8 standard with a gauge length of 25 mm and a width of 6 mm, termed A25.

\subsubsection{Preprocessing of microstructural images from experimental validation}

Each heat treatment process yielded 4 SEM images, totaling 24. We selected four experimental images, obtained at an annealing temperature of 815 $ \rm^\circ C $ and an annealing time of 3 min, as the validation results for DP1180. In addition, experimental images obtained at annealing temperatures of 765 $ \rm^\circ C $ and 715 $ \rm^\circ C $, and an annealing time of 13 min, were chosen as the validation results for DP980 and DP780, respectively.

To facilitate comparison with the design results, the microstructural images of the experimental steels required preprocessing. Initially, the experimental images were binarized using the ImageJ software~\cite{RN262}. Based on the average scale of all the images in the dataset, the experimental images with a resolution of 1024×768 pixels were resized to 833×720 pixels to ensure the scale of experimental images was close to that of literature-collected images. Subsequently, all the experimental images were sequentially sampled with a sampling size of 224×224 pixels, resulting in 108 sub-images (36 sub-images each for DP780, DP980, and DP1180). Upon feeding these images into the VAE model, the corresponding latent vectors and generated images were computed.

\section{Results and discussion}
\subsection{Construction of CPSP relationship based on VAE–DLM}

\begin{figure}[H]
\centering
\includegraphics[width=6.5in]{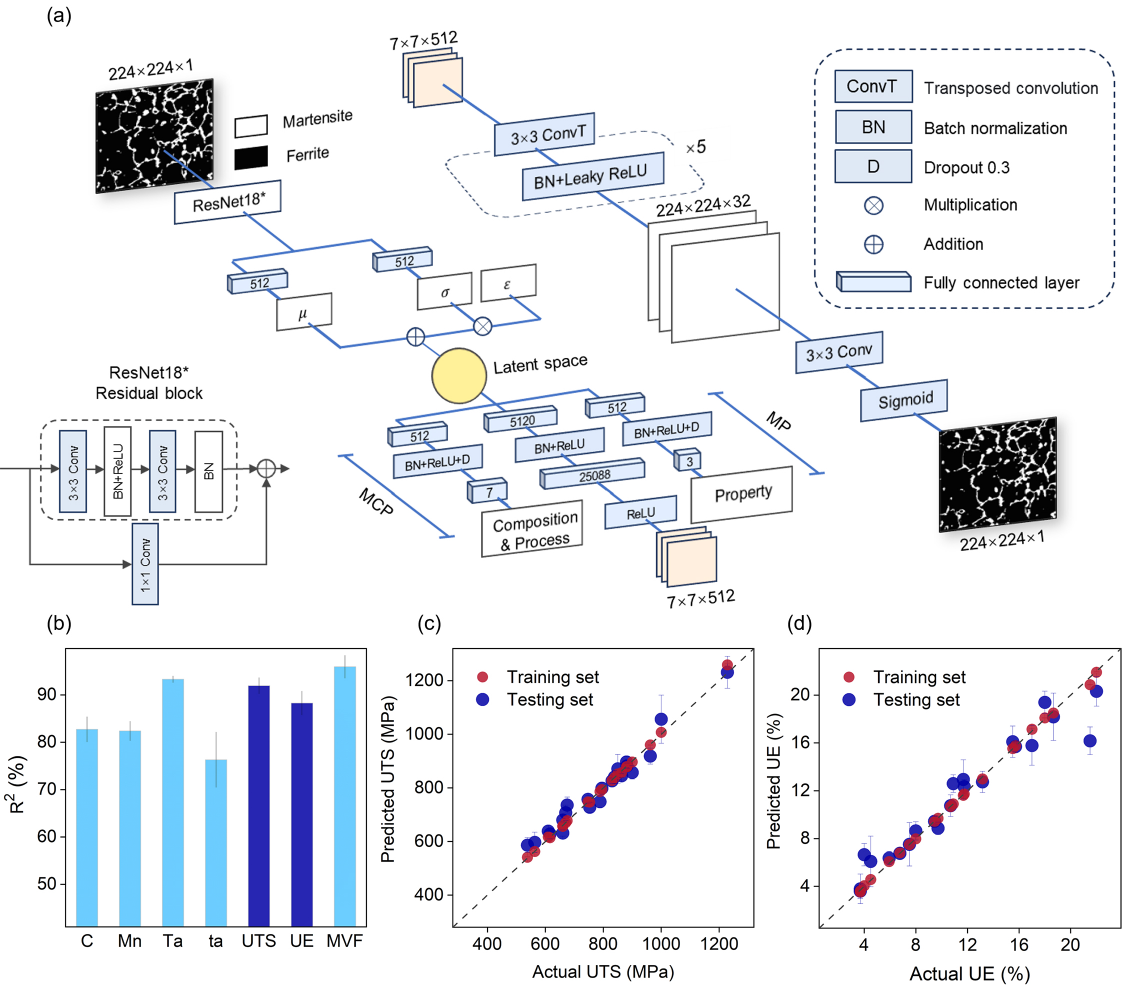}
\caption{\textbf{Architecture and predictive capabilities of the 
VAE-centric deep learning model (VAE–DLM).} (a) Model structure including an encoder, a decoder, and two multilayer perceptrons (called the MCP and MP models). The encoder, adapted from ResNet18, processes binary images (ferrite in black, martensite in white) to produce a probability distribution characterized by the mean ($\mu$) and standard deviation ($\sigma$). The decoder reconstructs a similar image from a latent vector drawn from this distribution. Utilizing the latent vectors, the MCP model predicts the composition and process of the alloy while the MP model predicts its properties. (b) Predictive performance across all the outputs. The term “Ta” and “ta” denote the annealing temperature and annealing time. Comparison of predicted and actual values of (c) ultimate tensile strength (UTS) and (d) uniform elongation (UE), respectively. }
\label{fig: architecture}
\end{figure}

The fundamental principle of the deep learning model applied to build the CPSP relationship, termed VAE–DLM (Fig.~\ref{fig: architecture}a), is to condense microstructural images into a low-dimensional representation in the latent space, which is then used for information fusion and alloy design. Specifically, the model integrates three key elements: a VAE and two MLP models.

The VAE, which is a generative model renowned for its latent representation learning, comprises an encoder and a decoder. The encoder employs a modified ResNet-18~\cite{RN200} architecture, tailored to process grayscale images through a single-channel first layer. It maps the input data into a latent space, generating a mean and standard deviation that define a probability distribution. The decoder, which is a combination of transposed convolution, batch normalization, activation, and convolution layers, reconstructs the original input by sampling from the probability distribution of the encoder. The transposed convolutional layers of the decoder utilize a kernel size of 3×3, a stride of 2×2, and the same padding to preserve the spatial dimensions of the input. Both the MCP and MP models were composed of fully connected, batch normalized, activation and dropout layers. Each hidden layer has 512 neurons. Although the inputs of both models are vectors in the latent space, their outputs were distinctly different. The MCP model predicts the parameters related to the material composition and processing: carbon (C), manganese (Mn), intercritical annealing temperature, annealing time, and process type. In contrast, the MP model focuses on predicting the material properties: the UTS, UE, and MVF.

The predictions of VAE–DLM are shown in Fig.~\ref{fig: architecture}b-d. $R^{2}$ for most of the test set outputs surpassed 80$\%$, suggesting that the VAE–DLM had robust generalization capabilities and a high predictive accuracy. Considering that the VAE–DLM will be subsequently used for the alloy design of DP steels and that predicting the mechanical properties is a crucial step in performance-oriented alloy design, we further examined the prediction effects of UTS and UE, as shown in Fig.~\ref{fig: architecture}c, d. For the training set, nearly all the results aligned precisely on the line with a slope of 1 and showed small error bars. For the test set, most of the points were near the diagonal, although a few points in the high-UE region deviated. The MAEs for the UTS and UE were only 33.9 MPa and 1.35$\%$, respectively, suggesting that the models adequately learned the relationship between the microstructure and performance.

In addition, we also compared our methodology with two different series of ML models: the first series of ML models takes the composition and process as inputs~\cite{RN247,RN248}, whereas the second takes composition, process, and MVF~\cite{RN175,RN173}. The comparative analysis demonstrated that our approach, which uses microstructural images as input, outperforms these models in accurately predicting the material properties (Supplementary Figs.~\ref{fig: mean absolute error} and~\ref{fig: property prediction}). This highlights the importance of incorporating actual microstructural data and establishing complete CPSP relationships to enhance the prediction accuracy of the mechanical properties.

\subsection{UniDP steel design based on probability distribution sampling}

The CPSP relationship, as clarified by the VAE–DLM, offers novel perspectives on the alloy composition and process design of UniDP steels. By sampling the latent space, we can derive the composition, process parameters, and properties of the new DP alloys. In the context of the VAE model, the feature vector $\mathit{Z}$ in the latent space is contingent upon two important variables: the mean ($\mu$) and the standard deviation ($\sigma$):

\begin{equation}
Z = \mu +  \sigma \times \varepsilon                                      \end{equation}

\begin{figure}[H]
\centering
\includegraphics[width=6.5in]{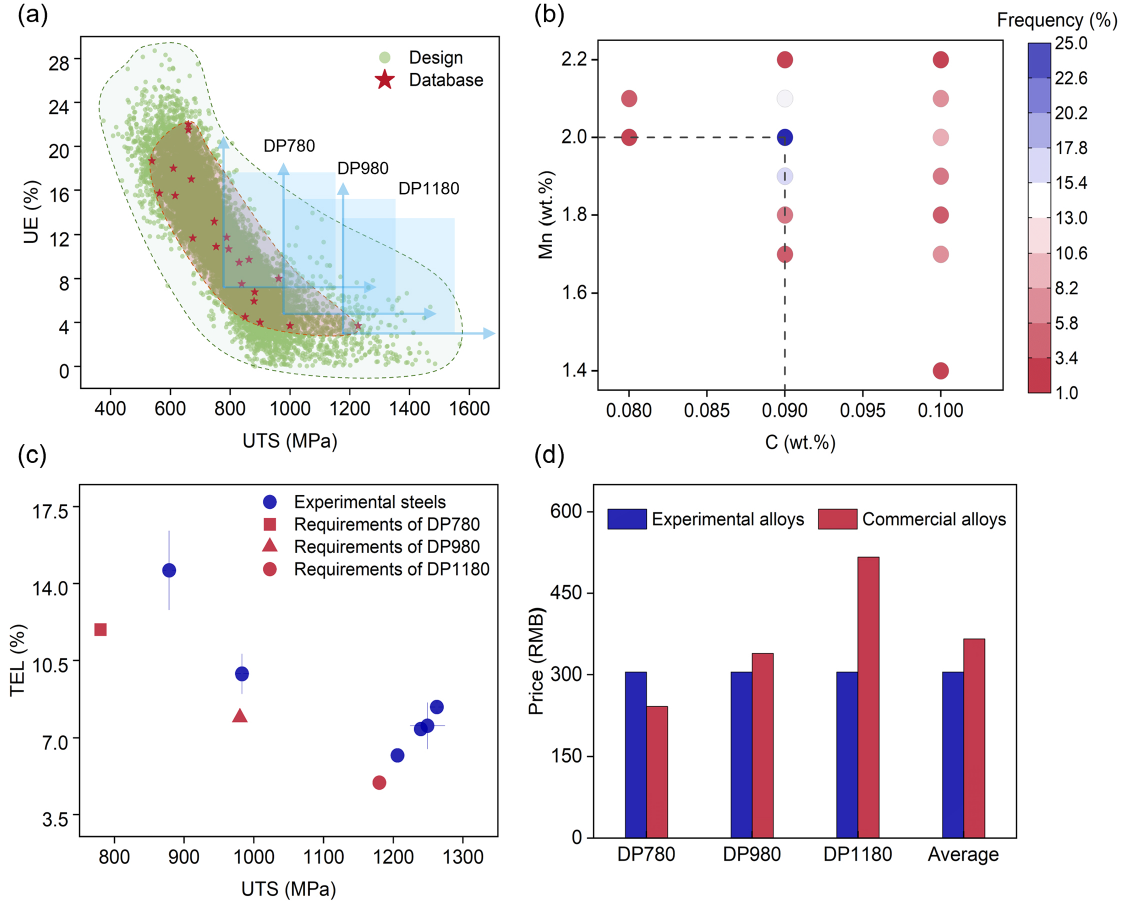}
\caption{\textbf{UniDP steel design.} (a) Mechanical property of the sampling points at $\varepsilon$ $\sim$ $\mathcal{N}$(0, $10^{2}$). (b) Compositional distribution of the alloy in the blue box in (a). (c) Mechanical properties of the experimental alloys. Designed new alloys meet the performance requirements of DP780, DP980, and DP1180, respectively. (d) Elemental cost comparison between the experimental and commercial alloys.}
\label{fig: design}
\end{figure}

\noindent
Where $\varepsilon$ adheres to a standard normal distribution during model training, denoted by $\varepsilon$ $\sim$ $\mathcal{N}$(0, 1). The 704 sub-images corresponded to the 704 $\mathit{Z}$ vectors. To investigate the impact of varying sampling ranges on the sampling results (refer to Supplementary Fig.~\ref{fig: trends}), the standard deviation of $\varepsilon$ was incrementally increased from 1 to 15. To ensure the validity of the results, each range was independently sampled 14,080 times, equating to 20 samples for each $\mathit{Z}$ vector. The findings indicate that a higher standard deviation yields more design outcomes that satisfy the DP980 performance criteria (UE and UTS) but also leads to an increase in the unreasonable design results (i.e., negative values of the composition, process, or properties, or predicted martensite volume fractions exceeding 100$\%$). Considering that a standard deviation of 10 for $\varepsilon$ preserves an accuracy of over 90$\%$ (reasonable sample proportion) and covers a wide range of properties (as depicted in Fig.~\ref{fig: design}a and Supplementary Fig.~\ref{fig: trends}), we chose this sampling result for this study and excluded any new unreasonable alloys.

In Fig.~\ref{fig: design}a, the design alloys that meet the performance requirements of DP780, DP980, and DP1180, hereinafter referred to as the initially screened alloys, are denoted by blue boxes. These performance criteria require UTS values exceeding 780, 980, and 1180 MPa, and corresponding UE values surpassing 7.21$\%$, 4.81$\%$, and 3.01$\%$, respectively. Figure~\ref{fig: design}b shows the compositional distribution of the initially screened alloys. The compositions of these alloys are primarily concentrated around C = 0.09 wt.$\%$ and Mn = 2.0 wt.$\%$, suggesting this composition as a reference for the composition of our UniDP steel. It is widely recognized that DP steels typically belong to the C–Mn–silicon (Si) system (referring to the GB/T 20564.2-2017 standard). Hence, based on the distribution characteristics of Si in the training dataset collected from literature, we adopted a mode value of Si = 0.42 wt.$\%$. For the heat treatment, we opted for quenching and intercritical annealing. This decision was informed by its frequent occurrence within the dataset and relative cost-effectiveness. The initially screened alloys (called secondary-screened alloys) that did not meet the above criteria for composition and process type were excluded. The initially screened alloys that meet the criteria (called candidate alloys) are used to determine the annealing temperature range for UniDP steels (refer to Supplementary Table~\ref{designed steels}). The annealing time was not designed because of the limited diversity within the training dataset.

Figure ~\ref{fig: design}c presents the experimental mechanical properties of the designed UniDP steels. All the designed UniDP steels had the same chemical composition with C = 0.09 wt.$\%$, Mn = 2.0 wt.$\%$, and Si = 0.42 wt.$\%$. Quenching and intercritical annealing were performed. The properties of DP780 and DP980 were attained at an annealing time of 13 min and annealing temperatures of 715 $ \rm^\circ C $ and 765 $ \rm^\circ C $, respectively. The properties of DP1180 were attained at an annealing temperature of 815 $ \rm^\circ C $ and annealing times of 3, 6, 9, and 13 min. An alteration in the strength level corresponds to microstructural transformation, particularly in the MVF (refer to Supplementary Figs.~\ref{fig: tensile curves} and~\ref{fig: SEM micrographs}). Figure~\ref{fig: design}d shows a cost comparison of the elemental constituents of the experimental steels and commercial steels. Because of the reduced Mn content, the designed UniDP steels exhibited lower elemental costs than commercial steels. Simultaneously, the singular heat treatment process flow of the experimental steels can help reduce the production expenses of enterprises. This highlights the significant potential inherent in current methodologies for designing new high-quality alloys.

\subsection{Comparison between candidate and experimental alloys}

\begin{figure}[H]
\centering
\includegraphics[width=6.5in]{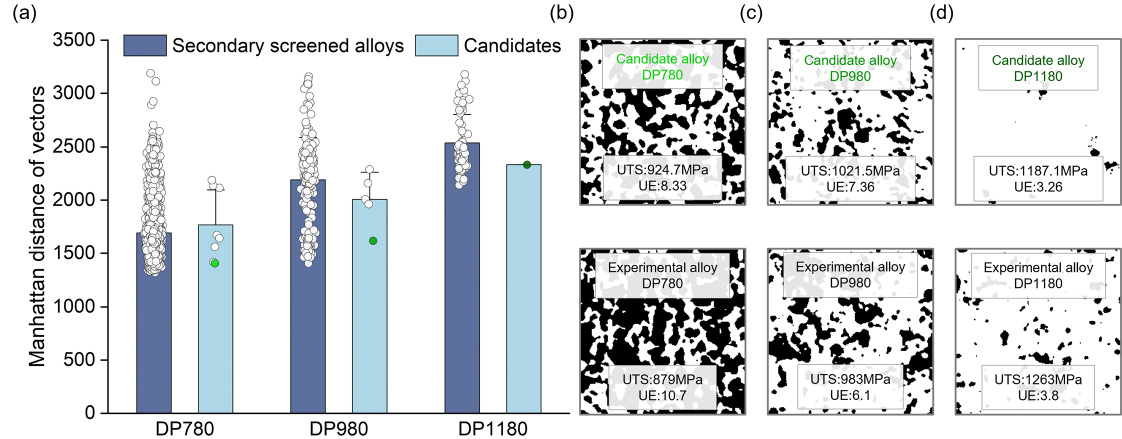}
\caption{\textbf{Comparison between candidate and experimental alloys.} Candidate alloys are novel alloys that have consistent composition and process as the experimental alloys and satisfy the performance criteria for different strength levels. Secondary-screened alloys are novel alloys that only satisfy the performance criteria for different strength levels. The encoder transforms the microstructural images of the experimental alloys into two vectors, namely mean $\mu$ and standard deviation $\sigma$. The mean $\mu$ is used as a latent vector for the experimental alloys, which is fed to the decoder to decode into a generated image of the experimental alloys. (a) Manhattan distances between the candidate/secondary-screened alloy and the experimental alloy vectors. The generated images and mechanical properties of the candidate alloys with the smallest vector distances are compared with those of the experimental alloys. Corresponding results for (b) DP780, (c) DP980, and (d) DP1180, respectively. The green fonts of different depths correspond to the vector points of the different candidate alloys in a), respectively.}
\label{fig: Comparison}
\end{figure}

To assess the rationality of our alloy design process, we compared the latent vectors, microstructural images generated, and mechanical properties of the experimental alloys and candidate alloys. The experimental images were preprocessed (refer to \textbf{Section 2.3.2}) and forwarded into the encoder. The Manhattan distances were then calculated between the latent vectors of the experimental alloys and those of the candidate/secondary-screened alloys. We used the mean value, denoted by $\mu$, of the encoder output as the latent vector for alloys. As depicted in Fig.~\ref{fig: Comparison}a, compared with the secondary-screened alloys, the average distances between the vectors of the candidate alloys and those of the experimental alloys were generally smaller. This suggests that the experimental alloys were more similar to the candidate alloys. Subsequently, we compared the generated images and properties of the candidate and experimental alloys with the smallest vector distances. As shown in Fig.~\ref{fig: Comparison}b–d, the generated images of both the sets of alloys exhibit numerous morphological similarities and comparable properties, with the average relative errors for the UTS and UE being 5$\%$ and 19$\%$, respectively. The experimental UniDP alloys resembled the candidate alloys closely, thereby providing a robust reference for UniDP steel design. This study integrated the actual microstructure, composition, and processing into the design, thereby effectively excluding unreasonable regions within the composition and processing space.

\subsection{Exploratory data analysis of latent space}

The latent vector space is formed by the probability distributions associated with all the images in the dataset. This space provided a qualitative framework for investigating the relationship between the attributes of alloys. The logic behind its construction directly influences the design results of the compositions, processes, and microstructures. Hence, it is necessary to visualize the latent space to reveal hidden correlations between attributes. In the following, we describe our methodologies and findings from the visualization.

\begin{figure}[H]
\centering
\includegraphics[width=6.5in]{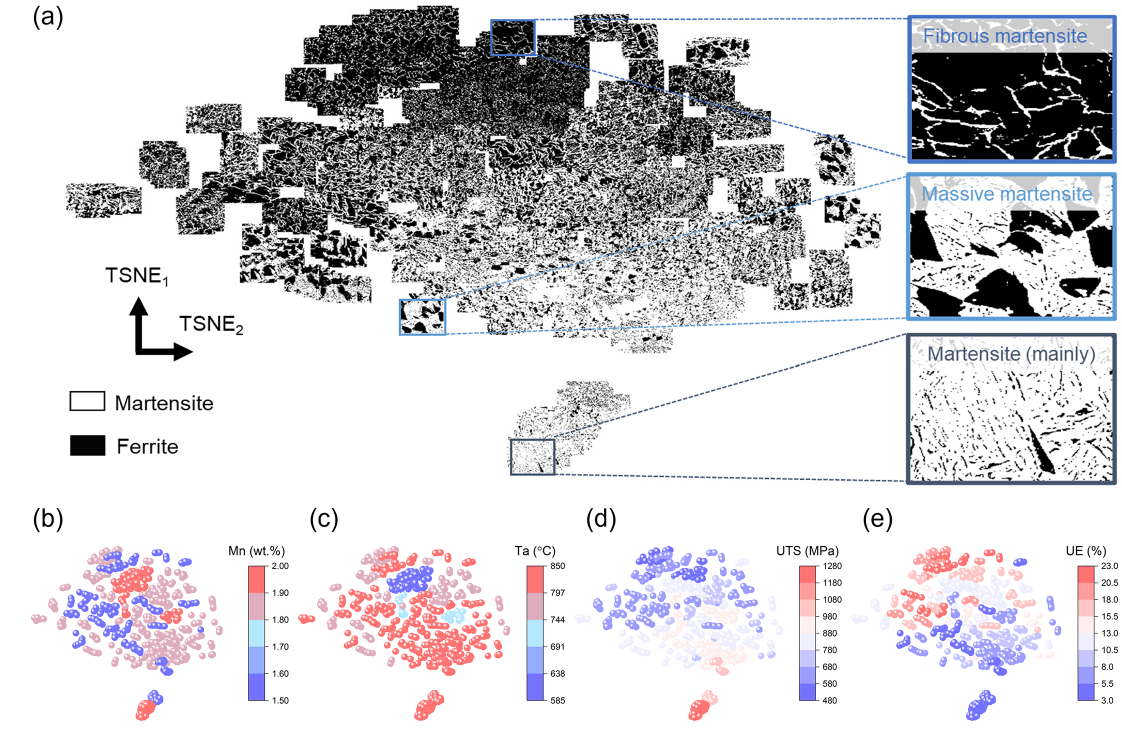}
\caption{\textbf{Visualization of the latent vector space.} The latent vector space comprises the probability distribution $\mathit{Z}$ of all the images in the dataset. To avoid sampling randomness, the probability distribution $\mathit{Z}$ is fixed to its mean $\mu$. Subsequently, the latent vector space is projected into the 2D space using t-SNE. The scatter points in the 2D space show the distributions of (a) microstructure, (b) manganese (Mn), (c) Ta, (d) UTS, and (e) UE. The term “Ta” denotes the annealing temperature.}
\label{fig: visualization}
\end{figure}

First, we utilized the mean, denoted by $\mu$, of the encoder output as the sampling result to mitigate randomness. The unsupervised t-SNE method~\cite{RN292} is then used to project the sampling vectors of the dataset images into a 2D space for visualization. Figure~\ref{fig: visualization}a shows the distribution of the microstructures following the t-SNE dimensionality reduction, with distinct morphologies being clearly delineated. From top left to bottom right, the microstructural images depict a progression. Initially, they present a combination of fibrous martensite and ferrite, followed by a mixture of blocky martensite and ferrite, and ultimately, they exhibit a structure composed mainly of martensite. These distinctly separated clusters suggest a robust correlation between the points in the t-SNE space and the microstructural images. This correlation was established within the constraints of the MP and MCP models, which encompass composition, process, and property information.

The t-SNE space illustrates the distribution of the Mn content, intercritical annealing temperature, and mechanical properties. As shown in Fig.~\ref{fig: visualization}b, c, most Mn elements and annealing temperature intervals showed a significant enrichment. The potential link between the two factors clearly demonstrates the characteristics of the dataset. However, the link between a single factor and the microstructure is more ambiguous. This could be due to the interplay of other factors, such as the annealing time and C content, highlighting the intricate relationships between the composition/process and microstructure and the challenges associated with adjusting the microstructure of DP steels using the conventional trial-and-error method. The UTS and UE distributions of the DP steels is shown in Fig.~\ref{fig: visualization}d, e. The clusters demonstrated a distinct pattern, with the UTS and UE progressively increasing and decreasing, respectively, from the upper left to the lower right. This pattern aligns with the previously mentioned microstructural variations and is in accordance with material science principles, indicating that the model could effectively comprehend the microstructural and property characteristics of DP steels.

\subsection{Continuous latent space}

A continuous interpolation of the latent vector space is crucial to material design frameworks based on the VAE–DLM. This is primarily because of its ability to generate a wide array of plausible new data through interpolation or out-of-domain sampling within a certain range. Hence, it is essential to investigate the continuity of latent vector spaces. We selected a sample image and fed it into the encoder to obtain the mean (denoted by $\mu$) and standard deviation (denoted by $\sigma$). Subsequently, we defined $\mathit{Z}$ = $\mu$ + $\tau$$\sigma$ and gradually increased the value of $\tau$. Finally, the latent vector $\mathit{Z}$ was fed into the decoder, MP, and MCP models to attain the generated image, UTS, and MVF, respectively.

\begin{figure}[H]
\centering
\includegraphics[width=6.5in]{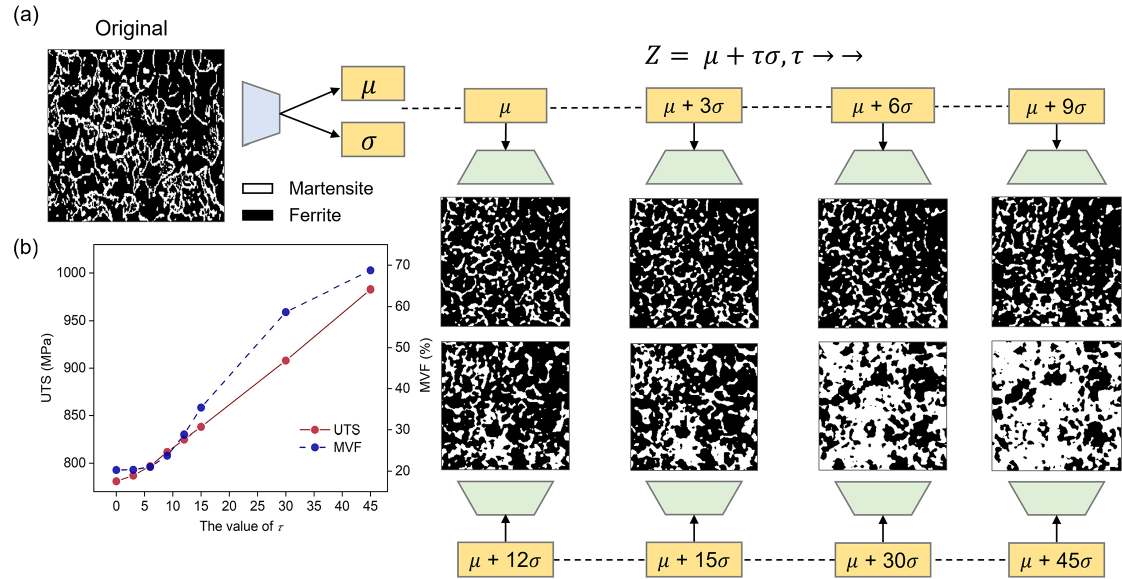}
\caption{\textbf{Continuously varying latent vector space.} The encoder transforms a microstructural image into two vectors: the mean $\mu$ and standard deviation $\sigma$. We set $\mathit{Z}$ = $\mu$ + $\tau$$\sigma$, with  as the variable, and convert $\mathit{Z}$ to a black-and-white image with a size of 224×224 through the decoder. Subsequently, $\mathit{Z}$ is inputted to the MCP and MP models to obtain the corresponding composition, process, and mechanical properties. Continuous variations in (a) microstructure and (b) UTS/ martensite volume fraction (MVF) when $\tau$ is varied from 0 to 45.}
\label{fig: space}
\end{figure}

Figure~\ref{fig: space} shows the alterations in the generated images, UTS, and MVF. The martensite phase (represented in white) within the generated image progressively expands toward the interior of the ferrite phase (depicted in black), transitioning from a fibrous to a blocky structure, and eventually coalescing into a structure predominantly composed of martensite. The corresponding MVF and UTS values increased incrementally. However, the growth rate of the MVF decelerated, while the growth rate of the UTS remained relatively constant, which might be attributed to an increase in the elemental content. Consequently, in the alloy design process, the design results corresponding to larger $\tau$ values might not satisfy the actual requirements when considering the issue of elemental cost. This further illustrates the significance of an appropriate sampling range for the successful design of UniDP steel. In conclusion, the latent vector space is well constructed, continuously interpolated, and can comprehensively consider the DP steel composition, process, properties, and microstructural information, thereby contributing to the UniDP steel design.

\section{Conclusion}
The design of alloys, particularly for multiphase alloys, presents formidable challenges in establishing accurate and complete CPSP links. To address this issue, we introduced a specialized deep learning model, namely the VAE–DLM, specifically for multiphase alloy systems and developed a novel alloy design framework centered on the microstructure. The framework merges advanced deep learning techniques with PM knowledge to create precise and robust CPSP connections from limited datasets. Moreover, this framework facilitates the rapid development of multiphase alloys using specific sampling strategies and PM knowledge. 

The effectiveness of this design approach was demonstrated by the development of a new UniDP steel. We compared the experimental results of the UniDP steel with its design results in terms of latent vectors, generated microstructures, and performance, justifying the design results. Moreover, through visualization analysis, we founded that the latent space generated by the VAE–DLM is both continuous and richly informative, which can aid in comprehending the intricate relationships between the material parameters and in providing robust guidance for the design of multiphase alloys.

This study represents a breakthrough in multiphase alloy design, stressing the critical incorporation of actual microstructural details into the CPSP framework to achieve high-fidelity correlations. In the future, we plan to extend our dataset and refine deep learning model to address more complex compositions of multiphase alloys, including quenching and partitioning steels. Moreover, ongoing research into the interpretative aspects of the latent space will improve the credibility of the framework in the design of multiphase alloys.

\subsection*{CRediT authorship contribution statement}

\textbf{Xudong Ma}: Conceptualization, Data curation, Formal analysis, Investigation, Methodology, Validation, Writing – original draft, Writing – review and editing. \textbf{Yuqi Zhang}: Conceptualization, Formal analysis, Investigation, Methodology, Software, Writing – review $\&$ editing. \textbf{Chenchong Wang}: Conceptualization, Formal analysis, Project administration, Supervision, Validation, Visualization. \textbf{Ming Wang}: Formal analysis, Resources, Supervision, Writing – review $\&$ editing. \textbf{Mingxin Huang}: Formal analysis, Funding acquisition, Project administration, Visualization. \textbf{Wei Xu}: Formal analysis, Funding acquisition, Project administration, Supervision.

\subsection*{Declaration of Competing Interest}
The authors declare that they have no known competing financial interests or personal relationships that could have appeared to influence the work reported in this paper.

\subsection*{Data availability}
The data that support the findings of this study are available from the corresponding author upon reasonable request.

\subsection*{Code availability}
The codes are available from the corresponding author upon reasonable request.

\subsection*{Acknowledgements}

The research was supported by the National Key Research and Development Program of China (No. 2022YFB3707501), the National Natural Science Foundation of China (No. U22A20106 and No. 52304392). M.X. Huang acknowledged the support from Mainland-Hong Kong Joint Funding Scheme, Platform (MHP/064/20).

% \bibliography{reference}
% \bibliographystyle{bibstyle}
\subsection*{Appendix. Supplementary materials}
See supplementary materials for more details.

\newpage
%\subsection*{Supporting information}
\begin{center}
	Supplemental Materials for\\[4mm]
	%\begin{flushleft}
	{\Large\textbf{Creating a Microstructure Latent Space with Rich Material Information for Multiphase Alloy Design}}\\
	%\end{flushleft}
\end{center}
\normalsize{Xudong Ma$ ^a $, Yuqi Zhang$ ^a $, Chenchong Wang$ ^{a,~*} $, Ming Wang$ ^{b,~*}$, Mingxin Huang$ ^{b,~*}$, Wei Xu$ ^{a,~*}$\\[6bp]
\textit{$ ^{\textit{a}} $ State Key Laboratory of Rolling and Automation, Northeastern University, Shenyang, Liaoning 110819, China}\\[6bp]
\textit{$ ^{\textit{b}} $ Department of Mechanical Engineering, The University of Hong Kong, Hong Kong, China}\\[6bp]
* Corresponding Author}\\[6bp]

\large
\noindent{\textbf{This supplementary material includes:}}\\
\textbf{Supplementary Figure~\ref{fig: details of data}.} Details of data on dual-phase (DP) steels gathered from literature.\\
\textbf{Supplementary Figure~\ref{fig: mean absolute error}.} Mean absolute error for two series of machine learning models on ultimate tensile strength (UTS) and uniform elongation (UE).\\
\textbf{Supplementary Figure~\ref{fig: property prediction}.} Comparison of mechanical property prediction capabilities among existing methods, and the CPP and CPMP methods.\\
\textbf{Supplementary Figure~\ref{fig: trends}.} Trends in the number of alloys meeting DP980 performance requirements and the percentage of reasonable samples (accuracy), with an increasing standard deviation of the normal distribution $\varepsilon$.\\
\textbf{Supplementary Figure~\ref{fig: tensile curves}.} Tensile curves of alloys under varying intercritical annealing process parameters.\\
\textbf{Supplementary Figure~\ref{fig: SEM micrographs}.} SEM micrographs of samples of DP steels inter-critically annealed at different annealing temperatures and times.\\
\textbf{Supplementary Table~\ref{output distribution}.} Output distribution in the current dataset.\\
\textbf{Supplementary Table~\ref{designed steels}.}  Composition and intercritical annealing process details for various designed DP steels.\\
\textbf{Supplementary Table~\ref{experimental alloys}.} Composition and intercritical annealing process details for the experimental alloys.\\
\textbf{Supplementary Note 1}. Training strategy.\\
\textbf{Supplementary Reference}\\

\newpage
\renewcommand{\figurename}{Supplementary Figure}

\setcounter{figure}{0}

\begin{figure}[H]
\centering
\includegraphics[width=6.5in]{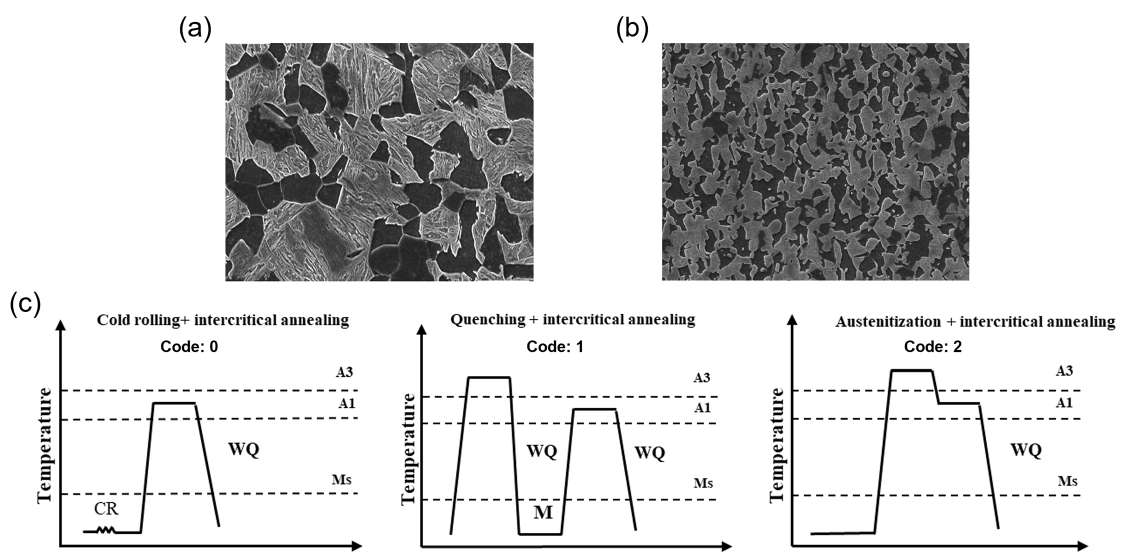}
\caption{Details of data on dual-phase (DP) steels gathered from literature. (a), (b) Representative scanning electron microscope (SEM) images. (c) Diverse heat treatment process routes. Different process routes correspond to different coding serial numbers to facilitate model training.}
\label{fig: details of data}
\end{figure}

\begin{figure}[H]
\centering
\includegraphics[width=6.5in]{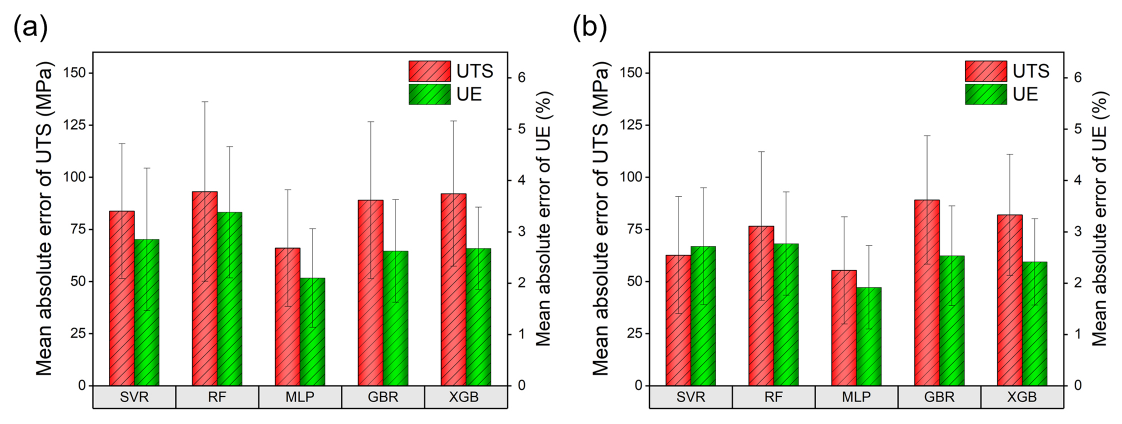}
\caption{Mean absolute error for two series of machine learning models on ultimate tensile strength (UTS) and uniform elongation (UE). (a) The first series (named CPP method) that uses composition and process as inputs. (b) The second series (named CPMP method) that uses composition, process, and martensite volume fraction as inputs. The machine learning models include support vector regression (SVR) [1], random forest (RF) [2], multilayer perceptron (MLP) [3], gradient boosting regression (GBR) [4] and extreme gradient boosting (XGB) [5]. Notably, MLP exhibits the lowest mean absolute error in both series, respectively.}
\label{fig: mean absolute error}
\end{figure}

\begin{figure}[H]
\centering
\includegraphics[width=6.5in]{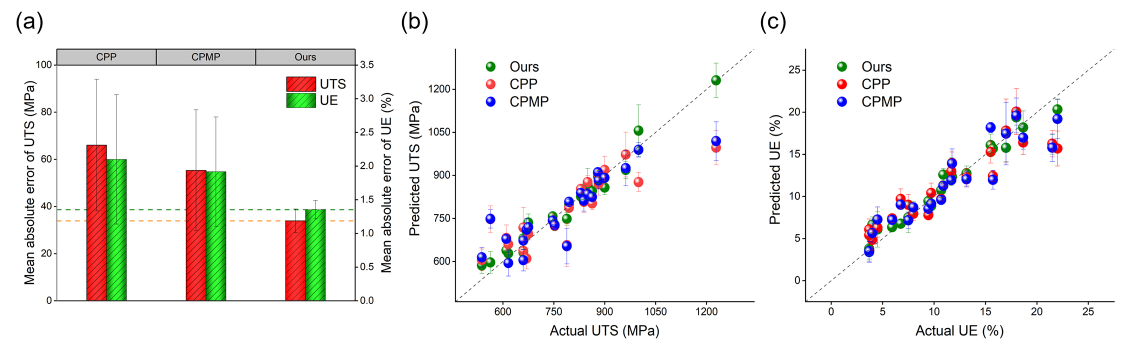}
\caption{Comparison of mechanical property prediction capabilities among existing methods, and the CPP and CPMP methods. (a) Mean absolute error across different methods. Predicted (b) UTS and (c) UE results from different methods. In this context, the results of the CPP and CPMP methods are generated from their respective MLP models. Despite this, our method still exhibits the lowest average absolute error, with predicted values closely aligning with target values in regions of high UE and UTS values.}
\label{fig: property prediction}
\end{figure}

\begin{figure}[H]
\centering
\includegraphics[width=6.5in]{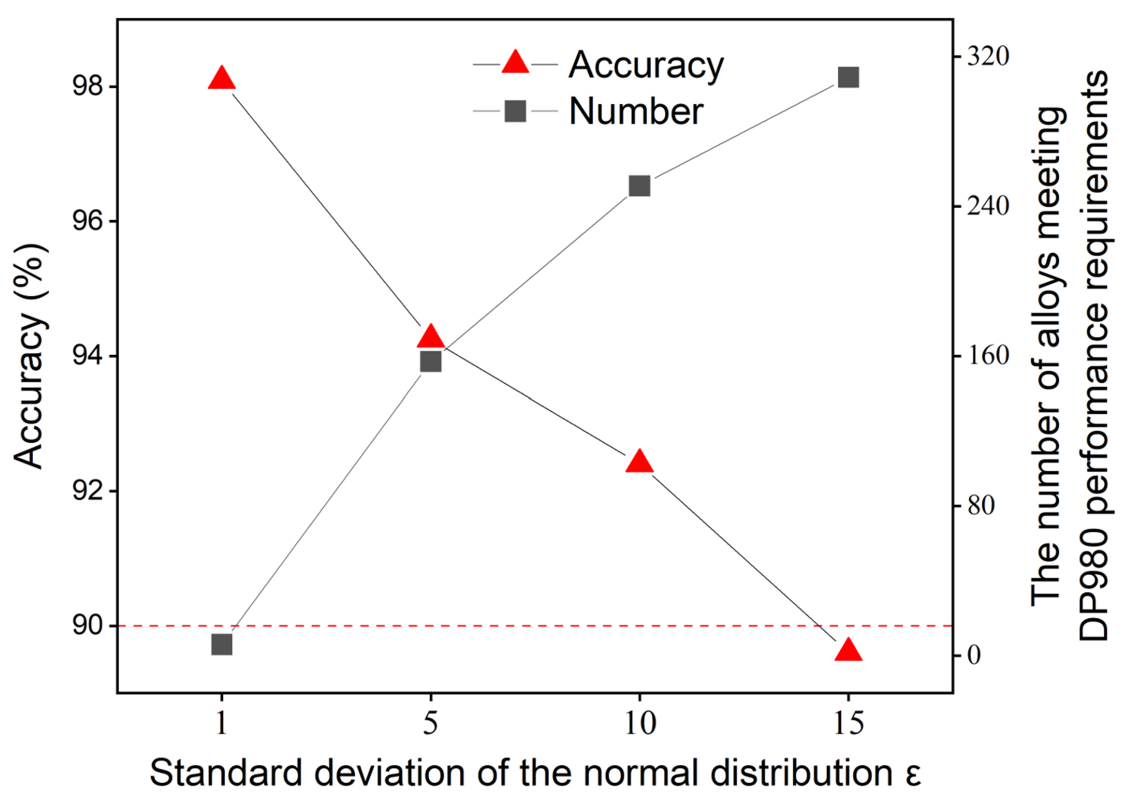}
\caption{Trends in the number of alloys meeting DP980 performance requirements and the percentage of reasonable samples (accuracy), with an increasing standard deviation of the normal distribution $\varepsilon$. In the context of the variational autocoder, the feature vector $\mathit{Z}$ in the latent space is influenced by two key variables, namely, the mean ($\mu$) and the standard deviation ($\sigma$), where $\mathit{Z}$ = $\mu$ +  $\sigma$ $\times$ $\varepsilon$. Here, $\varepsilon$ follows the standard normal distribution, $\varepsilon$ $\sim$ $\mathcal{N}$(0, 1), during model training. To investigate the effect of different sampling ranges on the sampling results, we gradually increase the standard deviation of the normal distribution $\varepsilon$ from 1 to 15 and sample separately. This process might yield samples with negative mechanical properties, compositions, and processes, or a martensite volume fraction greater than 100$\%$, which we term as unreasonable samples, and vice versa. Additionally, we use the number of design alloys that meet the DP980 performance criteria (refer to the Methods section for detailed information on performance criteria) as an evaluation metric for the design performance range. We posit that a greater number of alloys meeting the requirements indicates a broader design performance range.}
\label{fig: trends}
\end{figure}

\begin{figure}[H]
\centering
\includegraphics[width=6.5in]{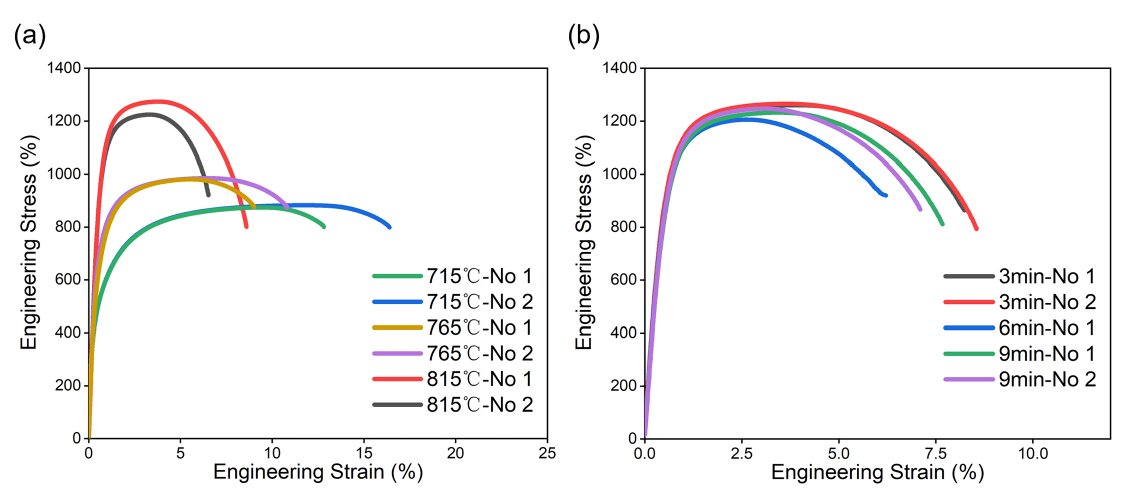}
\caption{Tensile curves of alloys under varying intercritical annealing process parameters. (a) Varying annealing temperatures with a constant annealing time of 13 min. (b) Varying annealing times at a constant annealing temperature of 815 $ \rm^\circ C $. Each experiment utilized two tensile specimens, with the exception of the experiment conducted at an annealing temperature of 815 $ \rm^\circ C $ and an annealing time of 6 min.}
\label{fig: tensile curves}
\end{figure}

\begin{figure}[H]
\centering
\includegraphics[width=6.5in]{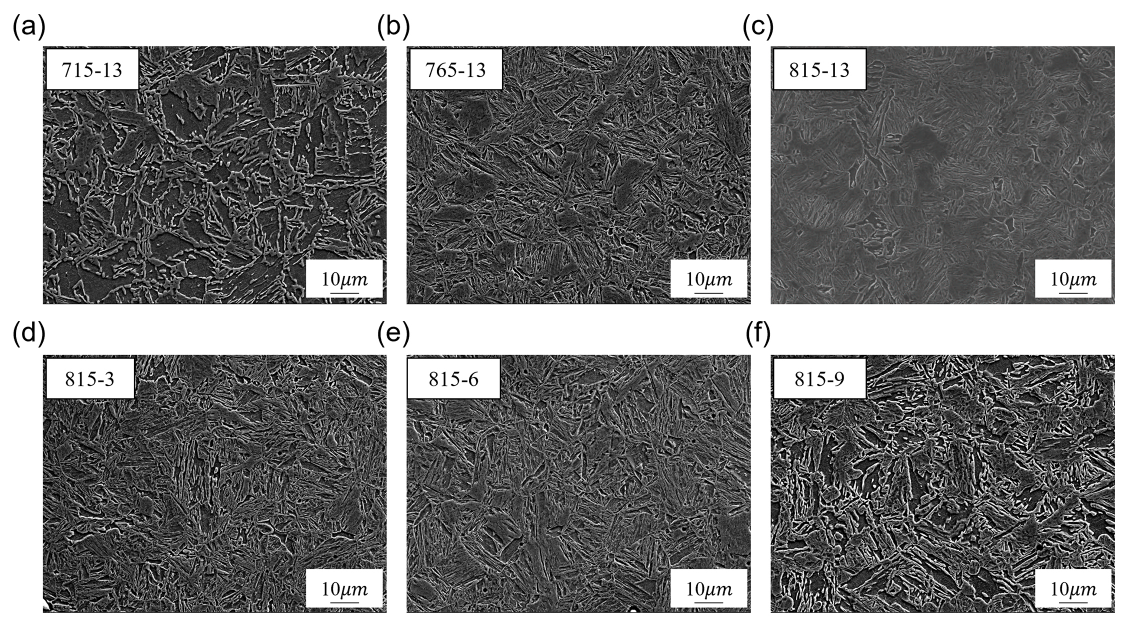}
\caption{SEM micrographs of samples of DP steels inter-critically annealed at different annealing temperatures and times: (a) 715$ \rm^\circ C $ and 13min; (b) 765$ \rm^\circ C $ and 13min; (c) 815$ \rm^\circ C $ and 13min; (d) 815$ \rm^\circ C $ and 3min; (e) 815$ \rm^\circ C $ and 6min; (f) 815$ \rm^\circ C $ and 9min.}
\label{fig: SEM micrographs}
\end{figure}

\renewcommand{\tablename}{Supplementary Table}
\setcounter{table}{0}

\begin{table}[H]
    \centering
    \caption{Output distribution in the current dataset.}
    \resizebox{\textwidth}{!}{
    \begin{tabular}{cccccc}
    \hline
        ~ & Output & Minimum & Maximum & Mean & \makecell[c]{Standard \\ deviation}  \\ \hline
        & C (wt.\%) & 0.08 & 0.093 & 0.09 & 0.01 \\ 
        ~ & Mn (wt.\%) & 1.50 & 1.93 & 1.74 & 0.17 \\ 
        ~ \makecell[c]{Composition and \\ processes} & Si (wt.\%) & 0.25 & 0.88 & 0.45 & 0.22 \\ 
        ~ & Annealing temperature ($ \rm^\circ C $) & 585.00 & 850 & 785.95 & 69.04 \\ 
        ~ & Annealing time (min) & 2.00 & 60 & 11.82 & 17.28 \\ \hline
         & Ultimate tensile strength (MPa) & 538 & 1228.4 & 786.82 & 158.05 \\ 
        ~ Properties & Uniform elongation (\%) & 3.69 & 22 & 11.36 & 5.59 \\ 
        ~ & Martensite volume fraction (\%) & 13 & 88 & 50.28 & 23.05 \\ \hline

    \end{tabular}}
    \label{output distribution}
    
\end{table}

\begin{table}[H]
    \centering
    \caption{Composition and intercritical annealing process details for various designed DP steels.}
    \resizebox{\textwidth}{!}{
    \begin{tabular}{cccccc}
    \hline
        ~ & C/wt.\% & Si/wt.\% & Mn/wt.\% & \makecell[c]{Annealing \\ temperature/$\rm^\circ C $}  & Process category \\ \hline
        Design DP780 & 0.09 & - & 2.0 & 703, 737, 753$…$ & \multirow{4}{*}{\makecell[c]{Quenching + intercritical  \\ annealing}} \\ 
        Design DP980 & 0.09 & - & 2.0 & 747, 814, 839 & ~ \\ 
        Design DP1180 & 0.09 & - & 2.0 & 823 & ~ \\ \cline{1-5}
        UniDP steel & 0.09 & 0.42 & 2.0 & 703 - 839 & ~ \\ \hline
    \end{tabular}
    }
    \label{designed steels}
\end{table}

\begin{table}[H]
    \centering
    \caption{Composition and intercritical annealing process details for the experimental alloys.}
    \resizebox{\textwidth}{!}{
    \begin{tabular}{ccccccc}
    \hline
    \large 
        ~ & C/wt.\% & Si/wt.\% & Mn/wt.\% & \makecell[c]{Annealing \\ temperature/$ \rm^\circ C $} & \makecell[c]{Annealing \\ time/min} & Process category  \\ \hline
        \multirow{2}{*}{\makecell[c]{Experimental \\ alloys}
        } & \multirow{2}{*}{0.096} & \multirow{2}{*}{0.42} & \multirow{2}{*}{1.97} & 715, 765, 815 & 13.0 & \multirow{2}{*}{\makecell[c]{Quenching +\\ intercritical annealing}}  \\ 
        ~ & ~ & ~ & ~ & 815 & 3.0, 6.0, 9.0 &   \\ \hline
    \end{tabular}}
    \label{experimental alloys}
\end{table}

\noindent
{\large\textbf{Supplementary Note 1}. Training strategy.}

\noindent
{\large The VAE–DLM has a total loss that can be expressed as follows:} 

\setcounter{equation}{0}

\begin{equation}
L = L_{KL} + L_{CE} + 300\left ( L_{CPP} + L_{CP} \right )
\end{equation}

\noindent
{\large where $L_{KL}$ represents the Kullback–Leibler divergence, and $L_{CE}$ denotes the cross-entropy loss. $L_{CPP}$ encompasses the regression losses of C, Mn, annealing temperature, annealing time, and classification loss of the process category, whereas $L_{CP}$ includes the regression losses of the UTS, UE, and MVF. The model was optimized over 3500 epochs, using a batch size of 64 within the PyTorch framework [6]. To achieve an optimal balance between the prediction accuracy and image reconstruction quality, we employed a step learning rate decay strategy with an initial learning rate of 0.001, coupled with the Adam optimization algorithm [7]. The learning rate decay strategy entailed a specific reduction in the learning rate to 0.9 times its current value after every 50 epochs.}

\noindent
{\large\textbf{Supplementary Reference}}

\begin{enumerate}[{[}1{]}]
\large
\item  J. Platt, Probabilistic outputs for support vector machines and comparisons to regularized likelihood methods, Advances in large margin classifiers. 10 (1999) 61-74. 
\item L. Breiman, Random Forests, Mach. Learn. 45 (2001) 5-32. 10.1023/A:1010933404324.
\item G.E. Hinton, Connectionist learning procedures, Artif. Intell. 40 (1989) 185-234. https://doi.org/10.1016/0004-3702(89)90049-0.
\item J.H. Friedman, Greedy Function Approximation: A Gradient Boosting Machine, Ann. Stat. 29 (2001) 1189-1232. 10.2307/2699986.
\item T. Chen, C. Guestrin, XGBoost: A Scalable Tree Boosting System,  Proceedings of the 22nd ACM SIGKDD International Conference on Knowledge Discovery and Data Mining, San Francisco, California, USA, 2016, pp. 785–794.
\item A. Paszke, S. Gross, F. Massa, A. Lerer, J. Bradbury, G. Chanan, T. Killeen, Z. Lin, N. Gimelshein, L. Antiga, A. Desmaison, A. Köpf, E. Yang, Z. DeVito, M. Raison, A. Tejani, S. Chilamkurthy, B. Steiner, L. Fang, S. Chintala, PyTorch: An Imperative Style, High-Performance Deep Learning Library, 2019.
\item D. Kingma, J. Ba, Adam: A Method for Stochastic Optimization,  International Conference on Learning Representations, 2014.
\end{enumerate}

\end{sloppypar}\end{document}